\documentstyle[12pt]{article}
\begin{document}

\begin{titlepage}

\hfill{October 1995}

\hfill{UM-P-95/94}

\hfill{RCHEP 95/21}

\vskip 1 cm

\centerline{{\large \bf
Maximal neutrino oscillation solution to the  }}
\centerline{{\large \bf
solar neutrino problem
}}

\vskip 1.5 cm

\centerline{R. Foot and R. R. Volkas}

\vskip 1.0 cm
\noindent
\centerline{{\it Research Centre for High Energy Physics,}}
\centerline{{\it School of Physics, University of Melbourne,}}
\centerline{{\it Parkville, 3052 Australia. }}

\vskip 1.0cm

\centerline{Abstract}
We discuss a simple predictive solution to
the solar neutrino problem based on maximal
vacuum neutrino oscillations.
The solution can be motivated by
the exact parity symmetric model
which predicts that the neutrino mass eigenstates
are maximal mixtures of ordinary and mirror weak eigenstates (if
neutrinos are massive).
We show that this proposed solution to the solar neutrino
problem is in
reasonable agreement with the experiments, and that in the
near future this scheme may be either ruled out or
tested more precisely as statistics improve for the
SAGE and GALLEX experiments.
Predictions are also given for the upcoming
Superkamiokande, SNO and Borexino experiments.

\vskip 1cm
\noindent

\end{titlepage}

The evidence for new neutrino physics from the comparison between
the measurements of the flux of solar neutrinos\cite{hom, kam,
gall, sage} and theoretical models of the sun\cite{x} has been
steadily accumulating.  Apparently strong evidence for a deficit
of atmospheric muon neutrinos has also emerged\cite{ana}.

An interesting possibility is that these discrepancies are
due to neutrino oscillations. For the solar neutrino
problem, a very popular model is based on the matter enhanced
MSW oscillations\cite{msw, xx}, which can significantly enhance the
conversion of electron neutrinos to another neutrino flavour for
a range of parameters.  This is an interesting idea, however the
main drawback of this possibility (in our opinion)
is its lack of predictivity. It can accomodate the data, but
it does not give any definite predictions for any of the
existing experiments (although given the existing
experimental results it does give significant
predictions for forthcoming experiments such as
SNO, Superkamiokande and Borexino).
For example, if a deficit had been
observed for the gallium experiments and not for the chlorine
experiment then this scenario could still have been interpreted
in terms of MSW oscillations with a
different range of parameters.\footnote{Also note
that if the LSND experiment\cite{lsnd} is correct
then the muon neutrino should have a mass of about 1 eV and hence
the MSW solution in terms of $\nu_e - \nu_{\mu}$ oscillations
is not possible.}

Another solution to the solar neutrino problem assumes
the existence of long wavelength vacuum  neutrino
oscillations (sometimes called ``just so'' neutrino oscillations)
\cite{js} which involves large angle vacuum oscillations
with oscillation length about equal to the distance between the
earth and the sun (which corresponds to $\delta m^2$ in the range
 $10^{-11} \stackrel{<}{\sim} \delta m^2/eV^2
\stackrel{<}{\sim} 10^{-10}$).
While this is an interesting possibility, we do not find it compelling
because of its lack of predictivity for any of the known
experiments and the limited range of parameters required. (In particular
it would be a lucky occurance if the distance of electron neutrino
oscillations just happened to be approximately the same as
the distance between the earth and the sun.)\footnote{
The MSW enhancement of solar neutrinos is also only available for
a limited range of parameters, but not as limited as
the long wavelength vacuum oscillation solution. Since solar
neutrinos (with energies relevant
for the experiments) can only encounter a
resonance for parameters in a limited range: $10^{-8}\ \stackrel
{>}{\sim}\ \delta m^2/eV^2\ \stackrel{>}{\sim}\ 10^{-4}$ and $\sin2\theta
\stackrel{>}{\sim}\ 10^{-2}$.}.

We would like to discuss an alternative solution to
the solar neutrino problem based on maximal neutrino
oscillations. We will focus on maximal
electron neutrino - sterile neutrino
oscillations (however our discussion will
be relevant to maximal electron neutrino - active neutrino oscillations
as well).
As is well known, for a large range of parameters
(i.e.  $10^{-10} eV^2 \stackrel{>}{\sim}
\delta m^2 \stackrel{>}{\sim} 10^{-3} \ eV^2$)\footnote{
Note that the $10^{-3}$ upper bound comes by requiring that
the electron neutrino does not oscillate over distances
relevant to the atmospheric neutrino experiment.
There is a laboratory bound of about $10^{-2}$ (see
Ref.\cite{pd})}
vacuum maximal oscillations implies that the flux
of electron neutrinos from the sun will be reduced by a factor of two
for all neutrino energies relevant to the solar
neutrino experiments.\footnote{
Note that if the oscillations are maximal then the flux of electron neutrinos
will be reduced by a factor of two independently of whether
there is any significant effect due to MSW oscillations
in the sun. This is because the vacuum neutrino oscillations
from the sun to the earth will mix the
electron neutrinos so that there are equal components of
electron neutrinos and sterile states.
However, if the electron neutrino mixes slightly with the
muon or tau neutrinos as well (and such mixing
would be expected)
then there could be additional effects due to MSW
oscillations due to this intergenerational mixing.
However, assuming that this mixing is small, the
region of parameter space where no significant effects occur
(e.g. $\delta m^2 \stackrel{>}{\sim} 10^{-4} \ eV^2$)
is much larger than
the region of parameter space where there is significant
enhancement of the neutrino fluxes.}
We will call this scenario the ``maximal vacuum oscillation
solution''. We believe this to be a very simple and
predictive scheme which can either be ruled out or tested more
stringently with the {\it existing experiments}.
Importantly, it also makes definite
predictions for the new experiments, SNO, Superkamiokande and
Borexino. [We stress that this solution is distinct from the
long wavelength  vacuum oscillation solution\cite{js}, which
has near maximal oscillations of electron neutrinos with another
species and $\delta m^2 \simeq 10^{-10}\
eV^2$. The main difference is that the long wavelength vacuum
oscillation  solution
is a best fit for the solar neutrino experiments (assuming
all these to be correct),
whereas the maximal oscillation solution
assumes that the mixing is exactly maximal and that
$\delta m^2 \stackrel{>}{\sim} 10^{-10}\ eV^2$ so that
the flux of solar neutrinos is reduced by a factor
of two and thus a definite prediction results.]
As far as we are aware, this simple idea
has never been discussed in detail, although it has been put
forward as a solution\cite{flv2, f, fv}.
Our interest in this scheme is motivated
by the exact parity symmetric model (see \cite{fv} for a
review of this model). This model
predicts that ordinary neutrinos will be
maximally mixed with mirror neutrinos (which are essentially sterile)
if neutrinos have mass\cite{flv2,f,fv}.\footnote{
It has been argued that the muon neutrino oscillation with a
sterile (mirror) neutrino with parameters $\delta m^2 = 10^{-2}\
eV^2, \sin^2 2\theta_0 \simeq 1$ (necessary to solve the atmospheric
neutrino anomaly) is inconsistent with standard Big
Bang Nucleosynthesis.
This is because oscillations with these parameters should
bring the sterile neutrino into equilibrium with the known neutrinos
\cite{B}. However this result depends critically on the assumption
that the relic neutrino asymmetries are small. Recently M.J.
Thomson and ourselves\cite{ftv}
have shown that large neutrino
asymmetries can be {\it created} by neutrino oscillations, and that
these bounds can be evaded.
This result can solve part of the potential cosmological
problem posed by the mirror particles. Note that one still
needs to postulate some new physics at high temperatures (or
a $t=0$ boundary condition) to bring about a
temperature difference between the mirror particles and the
ordinary particles.}$^,$\footnote{
Other models featuring maximal mixing have been discussed in the
literature within the context of neutrino anomalies. For instance,
Ref.\cite{giunti} discusses a model which has both electron and
muon neutrinos maximally mixed with two associated sterile species.
In addition they postulate that $\nu_e - \nu_{\mu}$ mixing occurs
in the MSW region. This last feature distinguishes their scenario
from ours. The papers in Ref.\cite{kobayashi} discuss a one-generation
model with maximally mixed electron and sterile neutrinos. These
works in addition invoke a large phenomenological neutrino magnetic
moment. Maximal mixing between electron neutrinos and other active
species has also been investigated (see for instance
Ref.\cite{joshipura}). Most of the above models use the idea of
pseudo-Dirac neutrinos \cite{wolfenstein}. Therefore, even those
models above which have maximally-mixed active-sterile systems
are distinct from the exact parity symmetric model from the point
of view of their construction. Our discussion of the solar and
atmospheric neutrino
problems will also be relevant to Ref.\cite{giunti}. Our discussion
of the solar neutrino problem only will be relevant
to those models which have $\nu_e$ maximally mixed with either $\nu_{\mu}$
or $\nu_{\tau}$.}
If we make the assumption that
mixing between generations is small (as it is in the quark sector)
then the parity symmetric model predicts that the three known neutrinos
will each be (to a good approximation) maximal mixtures
of two eigenstates.
This model thus nicely explains the atmospheric neutrino anomaly
which can be explained if the muon neutrino is maximally mixed
with a sterile neutrino (with $\delta m^2 \simeq 10^{-2} eV^2$)
\cite{f, fv, giunti}.

One nice feature of the parity symmetric model
is that its explanation of the
solar and atmospheric neutrino anomalies in terms
of maximal mixing between ordinary and mirror
neutrinos (which is a necessary consequence of the
parity symmetry) is that it is predictive and hence testible.
This is unlike other proposed explanations which involve
parameter fitting.  Another nice feature
is that it solves the atmospheric neutrino anomaly and
the solar neutrino deficit by essentially the same
mechanism: maximal vacuum oscillations. This
seems more appealing than the use of two different
mechanisms.

Despite the simplicity of the maximal mixing
solution to the solar neutrino deficit, no detailed
study of its comparison with existing experiments has
been made (so far as we know).
The purpose of this letter is to make such a study
using the most recent data.
We will show that the maximal mixing vacuum oscillation
solution is not excluded by existing experiments.
We will show that improved data from the existing gallium
experiments (SAGE and GALLEX) will be able to better test
or exclude the maximal mixing vacuum oscillation solution.

There are a number of theoretical calculations for the
solar neutrino flux. The uncertainties in the
theoretical calculations are largest for the $^8B$ solar
neutrino flux\footnote{For example, in the
model of Dar and Shaviv \cite{ds}, they obtain a theoretical
prediction for Kamiokande of $2.77 \times 10^6 cm^{-2}s^{-1}$ while in the
model of Bachall and Pinsonneault\cite{bp95} they
obtain a prediction of $6.6^{+0.9}_{-1.1} \times 10^6 cm^{-2}s^{-1}$.}.
However, for the flux of $pp, pep$ and $^7Be$
neutrinos there is good agreement among most (all?) theoretical
models. Thus instead of using the (possibly unreliable)
theoretical calculation for the  $^8B$ neutrino flux,
the empirical result from the Kamiokande experiment
can be used\cite{bb,kp}. To use this result in the other
experiments we need to assume that the shape of
the energy spectrum of boron neutrinos
as determined in laboratory experiments
with terrestrial sources is the
same as in the sun \cite{bahcall}. This will hold
unless some effect occurs to
distort the spectrum, like MSW oscillations \cite{bahcall}.
Maximal vacuum oscillations will not alter the spectrum.
It should thus be possible to make
reliable predictions for the other three experiments.

The most recent result of the
Kamiokande experiment\cite{kam} is
$$\phi_K(^8B) = (2.73\pm 0.17 \pm 0.34(syst))\times
10^6 cm^{-2} s^{-1}. \eqno (1)$$
The expected capture rate in the chlorine experiment
just from
$^8B$ neutrinos
(which we
denote as $R(^8B; ^{37}Cl)$)
as extrapolated from the Kamiokande experiment
(assuming the energy spectrum is unchanged) is
\cite{kp}
\footnote{
In Ref.\cite{kp}, they used the Kamiokande flux
measurement of $(2.89\pm 0.22 \pm 0.35(syst)) \times 10^6 cm^{-2} s^{-1}$,
and calculated the Chlorine capture rate of
$R(^8B, ^{37}Cl) \ge
2.94 \pm 0.40\ SNU$. However, the latest measurement of
the Kamiokande experiment has measured a flux slightly
smaller than the value used in Ref.\cite{kp}. Using this latest
value (Eq.(1)), we obtain the expected capture rate
for chlorine given in Eq.(2).
}
$$R(^8B; ^{37}Cl)  \ge 2.78 \pm 0.4\ SNU, \eqno (2)$$
where $1 SNU \equiv 10^{-36} $ captures per target atom per second.
Using the theoretical predictions for the other major
sources of neutrinos $^7Be$, $pep$, $pp$ and $CNO$, we obtain the
following expectations for the capture rates in the chlorine
and gallium experiments respectively\footnote{
For $^7Be$, $pep$ and $pp$ neutrinos we have used the values
listed in Ref.\cite{yy}, which have been obtained by
examining 10 different solar models. For $CNO$ neutrinos
we have taken a range of theoretical models
(from table 18 of Ref.\cite{tc})
and derived the error from the range of predictions.}:
$$R(^{37}Cl) = 4.5 \pm 0.5\ SNU $$
$$ R(^{71}Ga) = 123^{+8}_{-6}\ SNU \eqno (3)$$
where we have combined the various errors in quadrature
(we combine the errors in quadrature to get some idea of the total
error; this procedure is not strictly valid and the reader should
be aware that the true error may be different).
These predictions are summarised in table 1

\vskip 0.8cm

\tabskip=0pt \halign to \hsize{
\vrule#\tabskip=0pt plus 1fil\strut&
\hfil#\hfil& \vrule#&  \hfil#\hfil&
\tabskip=0pt\vrule#\cr
\noalign{\hrule}
&Chlorine&&Gallium&\cr
\noalign{\hrule}
&$^8B: 2.78\pm 0.4$&&$^8B: 7^{+7}_{-3.5}$&\cr
&$pep: 0.22\pm 0.01$&&$pp+pep: 74\pm 1$&\cr
&$^7Be: 1.1\pm 0.1$&&$^7Be: 34\pm 4$&\cr
&$CNO: 0.4\pm 0.2$&&$CNO:8.0 \pm 2.0$&\cr
&$Total: 4.5\pm 0.5$&&$Total:123^{+8}_{-6}  $&\cr
\noalign{\hrule}
}

\vskip 0.5cm
\noindent
Table 1: Expectation for the chlorine and gallium experiments using the $^8B$
flux as
extrapolated from the Kamiokande experiment and using the  theoretical
predictions from
a range of solar models for the other sources of neutrinos.
\vskip 0.9cm
\noindent
The experimental measurement for the chlorine
experiments is\cite{hom}\footnote{
Note that the average over the period
of operation (1970-1993) of the chlorine experiment
is $2.55 \pm 0.25 \ SNU$. However, some of the early
data has been criticised (see for example Ref.\cite{m}) on the
basis that there are quite significant fluctuations,
which are not apparent in the more recent Homestake data.
Also, since we are using the data from the Kamiokande
experiment as a measurement of the boron flux, it is
appropriate to use data collected at the same time.
For these reasons we use the data collected in the
period 1986-1993 and it is this data which
is given in Eq.(4).}
$$Cl: 2.78 \pm 0.35 \ SNU \eqno (4)$$
While for the two gallium experiments\cite{gall, sage} the
most recent measurements are
$$GALLEX:  77\pm 8(stat) \pm 5(syst)\ SNU;$$
$$SAGE: 69\pm 11(stat)\pm 6(syst) \ SNU. \eqno (5)$$
The weighted average of these two gallium measurements is
$75 \pm 9 \ SNU$ (where the systematic and statistical
errors have been combined in quadrature).
Comparing the theoretical predictions from table 1 with the
above experimental measurements, we see that there is
a very significant
discrepancy for both the chlorine experiment and the
gallium experiments.
While it is conceivable that the discrepancy with the chlorine
experiment could be due to some unaccounted systematic
error (for example in the neutrino-chlorine absorption cross section),
the discrepancy with the gallium predictions seems very robust.
(The radiodetection technique has been checked by
GALLEX by exposing their detector to a callibrated man-made
low energy neutrino source using neutrinos emitted in
the electron capture decay of ${51}Cr$\cite{check}).

How well does this maximal mixing vacuum oscillation solution
work for the solar neutrino problem?
If $\delta m^2 > 10^{-10} \ eV^2$ then the effect
of  maximal vacuum oscillations is simply to reduce the
predicted flux of all of the solar  neutrinos by a factor
of two. First we note that the Kamiokande experiment is in good
agreement with the predictions of most theoretical solar
models if the flux is halved.\footnote{For example, the 1995 analysis of
Bahcall and Pinsonneault\cite{bp95} predicts a
rate of $6.62^{+0.9}_{-1.1} \times 10^6 \ cm^{-2}
s^{-1}$ for Kamiokande. Dividing this by two
produces a rate of $3.31^{+0.45}_{-0.55}\times 10^6 \ cm^{-2}s^{-1}
$ which is in good agreement with the observed rate
Eq.(1). Note that for the case where $\nu_e$ is maximally mixed
with an active neutrino, one should divide the prediction by a
number slightly less than two (because Kamiokande is sensitive to
neutral currents). Good agreement is still obtained because of the
theoretical and experimental error ranges.}
However, as before, due to the larger uncertainties for $^8B$ neutrinos,
we will use the Kamiokande experiment
as a measurement of the $^8B$ flux of neutrinos, and use
the theoretical predictions for the fluxes of the other neutrinos.
The effect of the maximal mixing is to simply reduce
all of these predictions by a factor of two.
Thus, following the same steps as before but this time
dividing the theoretical fluxes for the $pp, pep, ^7Be,$ and $CNO$
neutrinos by a factor of two, we find the following
theoretical expectations for the fluxes assuming maximal
oscillations which we summarize in table 2: \footnote{Note
that this analysis also holds if $\nu_e$
is maximally mixed with an active species, because the chlorine
and gallium experiments cannot detect neutral current processes.
However the capture rate $R(^8B; ^{37}Cl)$ is a bit smaller
(about $2.4 \pm 0.5 \ SNU$)
because part (1/6 to 1/7) of the Kamiokande measurement would be due
to neutral current effects of the active $\nu_{\mu} $ or
$\nu_{\tau}$.}

\vskip 0.9cm

\tabskip=0pt \halign to \hsize{
\vrule#\tabskip=0pt plus 1fil\strut&
\hfil#\hfil& \vrule#&  \hfil#\hfil&
\tabskip=0pt\vrule#\cr
\noalign{\hrule}
&Chlorine&&Gallium&\cr
\noalign{\hrule}
&$^8B: 2.78\pm 0.4$&&$^8B: 7^{+7}_{-3.5}$&\cr
&$pep: 0.11\pm 0.005$&&$pp+pep: 37.0\pm 0.5$&\cr
&$^7Be: 0.55\pm 0.05$&&$^7Be: 17\pm 2$&\cr
&$CNO: 0.2\pm 0.1$&&$CNO:4.0 \pm 1.0$&\cr
&$Total: 3.64\pm 0.4$&&$Total:65^{+7}_{-4}$&\cr
\noalign{\hrule}
}

\vskip 0.5cm
\noindent
Table 2: Expectation for the chlorine and gallium experiments using the $^8B$
flux
as extrapolated from the Kamiokande experiment and the theoretical predictions
for the other fluxes assuming maximal vacuum oscillations (which have
the effect of halving the predicted fluxes).
\vskip 0.8cm
\noindent
We see that the prediction for the chlorine experiment is
significantly reduced by comparison with the minimal standard model
(see table 1).
The most dramatic effect occurs for the gallium experiment.
We summarise the results and compare with the experimental
measurements in table 3 below:
\vskip 0.7cm

\tabskip=0pt \halign to \hsize{
\vrule#\tabskip=0pt plus 1fil\strut&
\hfil#\hfil& \vrule#& \hfil#\hfil& \vrule#& \hfil#\hfil&
\tabskip=0pt\vrule#\cr
\noalign{\hrule}
&Prediction/Expt&&Chlorine&&Gallium &\cr
\noalign{\hrule}
&Standard Electro-weak theory&&$4.5\pm0.5$&&$123^{+8}_{-6}$&\cr
&Maximal mixing model&&$3.64
\pm 0.4 $&&$65^{+7}_{-4}$&\cr
&Experiment&&$2.78 \pm 0.35 $&&$75\pm 9$&\cr
\noalign{\hrule}
}

\vskip 0.5cm
\noindent
Table 3: Summary of the predictions for the chlorine and gallium experiments
assuming 1) standard electro-weak theory (i.e. no new physics) 2) that the
electron neutrino
oscillates maximally into a sterile state (maximal mixing model) and
3) the experimental measurements.
\vskip 0.8cm
\noindent
The prediction of the maximal mixing model for the gallium
experiments is obviously in excellent agreement with the data.
The discrepancy between the maximal
mixing prediction
for the chlorine experiment and the measurement is
$0.86\pm 0.4 \pm 0.35$ ($= 0.86\pm 0.53$ if the errors
are added in quadrature). This discrepancy
is only about one and a half $\sigma$. We therefore conclude that a
simple factor-of-two reduction of all the neutrino fluxes
(as predicted by the maximal mixing vacuum oscillation scheme) is
sufficient to bring about a reasonably good reconciliation
between theory and experiment.

The discrepancy with the chlorine
experiment is suggestive of energy
dependence to the solar neutrino deficit. Indeed,
the relatively low flux observed by Homestake is the prime reason
why an energy dependent reduction of neutrino fluxes (as
given by the MSW solution and ``just so'' vacuum oscillation
solution) has received much attention.
However, within the context of the maximal mixing vacuum
oscillation solution, the low Homestake result becomes
less than  a $2\sigma$ effect. A discrepancy of less than
2$\sigma$ in
an experiment like Homestake should not be considered
compelling evidence for an
energy dependence to the neutrino deficit. It could, instead,
be due to an unaccounted-for systematic error, or
simply to not enough statistics.

Note that our
prediction will become more precise as more data is collected
from Kamiokande. It will be interesting to see if this
discrepancy is reduced as this data is collected.
A more important test will be the gallium experiment.
At present there is good agreement between the prediction of
the maximal mixing solution and the measurements.
The error on the measurements will be reduced as more data
is collected. A final measurement above about 82 SNU would
make the maximal mixing vacuum oscillation solution an unlikely
explanation, while a final measurement below about 72 SNU
would strongly support it.

What does the maximal mixing solution to the solar neutrino
problem predict for the forthcoming experiments:
the Sudbury Neutrino Observatory (SNO) experiment,
Superkamiokande and Borexino?
Superkamiokande and SNO will be able to measure
the shape of the energy spectrum of the neutrinos
from $^8B$. The maximal vacuum oscillation solution predicts the same
energy spectrum as measured in the laboratory (MSW predicts
a significant deviation). The maximal mixing solution also predicts no
day/night effect. This effect could only potentially occur
if $\delta m^2 \sim 10^{-6} \ eV^2$. However, since any
regeneration of electron neutrinos in the earth
will be compensated by a depletion of electron neutrinos (as
there are equal amounts of $\nu_e$ and $\nu_s$ due to
the maximally mixed oscillations during the propagation
of neutrinos from the sun to the earth) there should be no significant
day/night effect if the maximal oscillation solution is correct.

SNO will also be able to distinguish neutral current interactions from
charged current interactions and
thus test whether neutrinos oscillate into active neutrinos
or sterile neutrinos. Thus the prediction of the maximal  oscillation
solution depends on whether the maximal
oscillations occur between active and sterile neutrinos
(as in the parity symmetric model) or between active and
active neutrinos. Thus the maximal mixing solution with neutrinos
oscillating maximally into sterile neutrinos predicts that
both neutral and charged currents will be reduced by
a factor of two. In other words the ratio of
charged to neutral current events will be the same
as the minimal standard model with no oscillations.

Superkamiokande and SNO will be able to search for seasonal effects.
Note that the maximal oscillation solution does not
predict any significant seasonal effect beyond that due
to the inverse square law (as distinct from
the long wavelength ``just so'' oscillation scenario).

Finally, Borexino will be able to give a precise measurement
of the $^7Be$ line. The maximal oscillation solution predicts a reduction
of $1/2$ compared to the standard model. This experiment would
be another good test of the maximal oscillation solution [as
the current expectation from the MSW and long wavelength (``just so'')
solutions is that there will be a much greater reduction
in the $^7Be$ signal].

The maximal oscillation solution to the solar neutrino problem
was motivated by the parity symmetric model. Even if
this solution turns out to be incorrect the parity
symmetric model may still be connected to the solar neutrino
 problem. For example the large wavelength vacuum oscillation
solution can be fit with maximal oscillations if $\delta m^2 \approx
10^{-10} \ eV^2$. Alternatively, one can have MSW type oscillations
between the electron neutrino and the muon neutrino together with
maximal oscillations of the electron neutrino and the sterile neutrino (
this scenario was discussed in the model of Ref.\cite{giunti}.).
However, in our opinion the maximal oscillation solution
is a more
likely solution.

\vskip 1.1cm
\centerline{\bf Acknowledgements}
\vskip .3cm
\noindent
This work was support by the Australian Research Council.

\end{document}